\documentclass{iau}
\usepackage{graphicx}
\usepackage{amsmath}
\usepackage{natbib}
\bibpunct{(}{)}{;}{a}{}{,} % to follow the A&A style

\newlength{\digit}
\newcommand{\dd}{{\hspace*{\digit}}} 
\newcommand{\tc}[1]{\multicolumn{1}{c}{{#1}}}
\newcommand\tstrut{\rule{0pt}{2.5ex}}

\newcommand{\NF}{{$N_\mathrm{frag}\/$ }}

\title[Fragmentation of colliding planetesimals with water content] %% give here short title %%
{Fragmentation of colliding planetesimals\\ with water content}

\author[T.~I.\ Maindl, R.\ Dvorak, C.\ Sch{\"a}fer \& R.\ Speith]
{Thomas I.\ Maindl$^1$,
 Rudolf Dvorak$^1$,
 Christoph Sch{\"a}fer$^2$
 \and Roland Speith$^3$}

\affiliation{$^1$Department of Astrophysics, University of Vienna,
              T{\"u}rkenschanzstra\ss e 17, A-1180 Vienna, Austria,
              email: {\tt thomas.maindl@univie.ac.at, rudolf.dvorak@univie.ac.at}\\[\affilskip]
$^2$Institut f{\"u}r Astronomie und Astrophysik, Eberhard Karls Universit{\"a}t T{\"u}bingen,
Auf der Morgenstelle 10, 72076 T{\"u}bingen, Germany,
             email: {\tt ch.schaefer@uni-tuebingen.de}\\[\affilskip]
$^3$Physikalisches Institut, Eberhard Karls Universit\"at T\"ubingen,
Auf der Morgenstelle 14, 72076 T\"ubingen, Germany,
             email: {\tt speith@pit.physik.uni-tuebingen.de}}

\pubyear{2014}
\volume{310}  %% insert here IAU Symposium No.
\pagerange{119--126}
\jname{Complex planetary systems}
\editors{A.C. Editor, B.D. Editor \& C.E. Editor, eds.}
\begin{document}

\maketitle

\begin{abstract}
We investigate the outcome of collisions of Ceres-sized planetesimals composed of a rocky core and a shell of water ice. These collisions are not only relevant for explaining the formation of planetary embryos in early planetary systems, but also provide insight into the formation of asteroid families and possible water transport via colliding small bodies. Earlier studies show characteristic collision velocities exceeding the bodies' mutual escape velocity which---along with the distribution of the impact angles---cover the collision outcome regimes %often called 
`partial accretion', `erosion', and `hit-and-run' leading to different expected fragmentation scenarios. Existing collision simulations use bodies composed of strengthless material; we study the distribution of fragments and their water contents considering the full elasto-plastic continuum mechanics equations also including brittle failure and fragmentation.

%% add here a maximum of 10 keywords, to be taken form the file <Keywords.txt>
\end{abstract}

\firstsection % if your document starts with a section,
              % remove some space above using this command.
\section{Introduction}

Most simulations of giant impacts use a strengthless material model \citep[e.g.,][]{canbar13} based on the fact that beyond a certain size \citep[400\,m in radius,][]{melrya97} self-gravity dominates the material's tensile strength. In \citet{maidvo13b} we study colliding objects of a size close to this limit and compare strengthless material simulation (``hydro model'') to a model including material strength leading to elasto-plastic effects and the possibility of brittle failure (``solid model''). These investigations are done with our own smoothed particle hydrodynamics (SPH) code as introduced in \citet{maisch13,maidvo13b} implementing the \citet{grakip80} fragmentation model \citep[see also][]{benasp94} and ensuring first-order consistency \citep[tensorial correction, described in][]{schspe07}. Our scenarios include two objects, the target consisting of a basalt core (70\,mass-\%) and a shell of water ice (30\,mass-\%) and the projectile consisting of solid basalt. Projectile and target have a mass of $M_\mathrm{Ceres}\/$ each. In total we simulate 42 scenarios (each in both the hydro and solid models) defined by collision velocities $v_0\/$ between 0.95 and 5.88 two-body escape velocities $v_\mathrm{esc}\/$ \citep[which were found representative for such impacts in][]{maidvo13} and impact angles $\alpha\/$ between $0^\circ\/$ (head-on) and a flyby. We find that while the collision outcome in terms of merging/erosion/hit-and-run for the solid model is similar to the hydro case we observe a significantly higher degree of fragmentation and more water loss in the solid case.

Here we will focus on analyzing ``significant fragments'' surviving the collision in the solid case regarding their number and water content after the impact as we are interested in water transport mechanisms in the early solar system \citep{dvoegg12}.

\section{Results}

SPH-based simulations resolve continuous bodies into discrete SPH particles carrying the physical properties of interest and contributing to the physical properties of their respective volume elements. Because of the high degree of fragmentation after impacts in the solid model and in order to get a statistically significant number of these discrete particles we limit our study to fragments with masses $m_\mathrm{frag}\/$ corresponding to at least 20 basalt particles. With our scenarios' single basalt SPH particle mass of $1.2\times 10^{17}\,\mathrm{kg}\/$ this translates to $m_\mathrm{frag}\ge 2.4\times 10^{18}\,\mathrm{kg}\/$ which limits the resolution of our results. Based on the 42 different collision configurations simulated in \citet{maidvo13b} we get the numbers \NF of such ``significant fragments'' as listed in Table~\ref{tab:nfrags}. Each $\alpha\/$-\NF column pair corresponds to one initial impact parameter. As projectile and target were placed five mean diameters apart at the start of the simulation the impact parameters and angles $\alpha\/$ changed during the approach depending on mutual gravitational interaction resulting in the values given in the $\alpha\/$ columns% \citep[for details, see][]{maidvo13b}
.
\begin{table}
\begin{center}
\begin{tabular}{r@{\qquad}*{5}{rc@{\qquad}}rc}
\hline \hline
\tstrut$v_0$ &	\tc{$\alpha$} & \NF & \tc{$\alpha$} & \NF & \tc{$\alpha$} & \NF & \tc{$\alpha$} & \NF & \tc{$\alpha$} & \NF & \tc{$\alpha$} & \NF \\
\tstrut$[v_\mathrm{esc}]$ & \tc{[$^\circ$]} & & \tc{[$^\circ$]} & & \tc{[$^\circ$]} & & \tc{[$^\circ$]} & & \tc{[$^\circ$]} & & \tc{[$^\circ$]} \\
\hline						
\tstrut
0.95 & 0 &\dd 1 & 14 &\dd 1 & 23 &\dd 1 & 25 &\dd 1 & 31 &\dd 1 & 62     &\dd 7 \\
1.32 & 0 &\dd 1 & 11 &\dd 1 & 21 &\dd 1 & 40 &\dd 1 & 48 &\dd 2 & \tc{-} &\dd 2 \\
1.36 & 0 &\dd 1 & 11 &\dd 1 & 20 &\dd 1 & 40 &\dd 2 & 48 &\dd 2 & \tc{-} &\dd 2  \\
2.12 & 0 &\dd 1 & 12 &\dd 1 & 25 &\dd 5 & 50 &\dd 2 & 62 &\dd 6 & \tc{-} &\dd 2  \\
3.04 & 0 &   39 & 12 &   52 & 25 &   22 & 53 &\dd 2 & 67 &\dd 2 & \tc{-} &\dd 2  \\
3.97 & 0 &   41 & 13 &   67 & 27 &   35 & 55 &\dd 2 & 72 &\dd 2 & \tc{-} &\dd 2  \\
5.88 & 0 &   45 & 13 &   60 & 28 &   61 & 58 &\dd 2 & 72 &\dd 2 & \tc{-} &\dd 2  \\
\hline

\end{tabular}
\end{center}
\caption{Number of significant fragments after 2000\,min simulation time. The scenarios are characterized by the collision velocity $v_0\/$ given in units of the two-body escape velocity upon impact $v_\mathrm{esc}$ and the impact angle $\alpha\/$. The $v_0\/$ (averaged) and $\alpha\/$ values are taken from \citet{maidvo13b}.\label{tab:nfrags}}
\end{table}

\subsection{Fragment properties}

\begin{figure}
% \vspace*{-2.0 cm}
\begin{center}
 \includegraphics[width=0.9\textwidth]{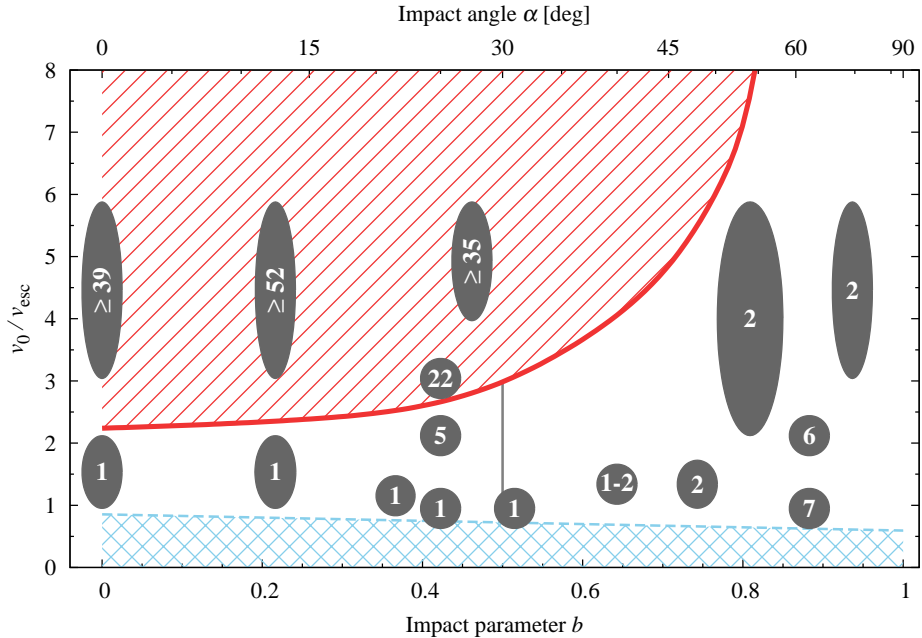} 
% \vspace*{-1.0 cm}
 \caption{Significant surviving fragments for different impact angles $\alpha\/$ (corresponding to a dimensionless impact parameter $b=\sin\alpha\/$) and collision velocities $v_0\/$ in units of the two-body escape velocity $v_\mathrm{esc}\/$. The different areas correspond to erosion (red, shaded), partial accretion (white, $\alpha \lesssim 30^\circ\/$), hit-and-run (white, $\alpha \gtrsim 30^\circ\/$), and perfect merging (blue, hashed) given by the analytic model for strengthless planet collisions in \citet{leiste12}, Fig.~11A. See text for discussion.}
   \label{fig:collmap}
\end{center}
\end{figure}

In Fig.~\ref{fig:collmap} we relate the numerical results to an analytic model for the outcome of strengthless planet collisions as presented by \citet{leiste12}. In the erosion regime we find many surviving fragments, which is consistent with the analytic prediction. Also, most of the one-fragment and two-fragment outcomes are in the appropriate regions. There is indication however, that the onset of hit-and-run events happens at a higher impact angle than predicted in the analytic strengthless model: while at an impact angle of $40^\circ\/$ a velocity of $v_0=1.32\,v_\mathrm{esc}\/$ leads to a merge, a slightly higher $v_0=1.36\,v_\mathrm{esc}\/$ produces a hit-and-run outcome (cf.\ the ``1--2'' bubble in Fig.~\ref{fig:collmap}). At $\alpha=25^\circ\/$ there is a \NF$=5$ outcome that suggests the border between merging and erosion to actually be at lower velocities for colliding smaller bodies. A closer look at that particular scenario reveals two major surviving fragments amounting for about 80\,\% of the system mass and several smaller ones. At $\alpha=62^\circ\/$, the \NF$=6$ outcome corresponds to a hit-and-run where the two main survivors amount for not quite the total mass of the system (around 90\,mass-\%) and several minor fragments. The \NF$=7$ case however marks the border to the merging area: among the surviving significant fragments there is only one major survivor ($\ge 90$\,mass-\%). For more details see the systematic discussion of the collision outcomes using a three-biggest-fragment approach in \citet{maidvo13b}.

\subsection{Water content}

\begin{figure}
% \vspace*{-2.0 cm}
\begin{center}
 \includegraphics[width=0.9\textwidth]{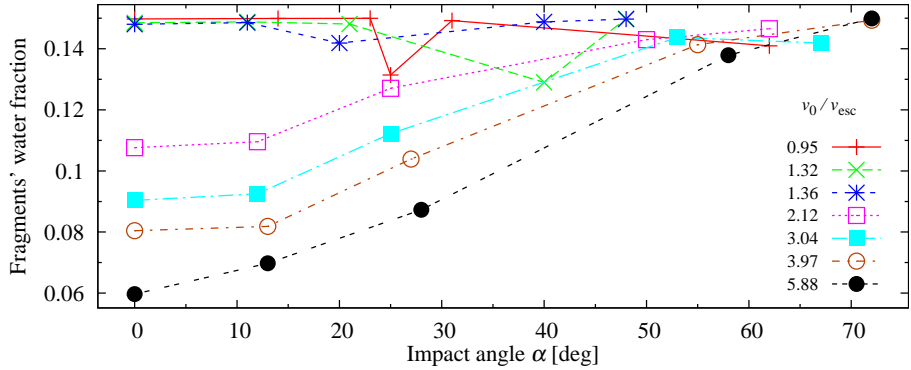} 
% \vspace*{-1.0 cm}
 \caption{Cumulated water content of significant fragments after the collision (mass fraction). The total water in the system is 15\,mass-\%. The different curves correspond to different collision velocities $v_0\/$; $v_\mathrm{esc}\/$ is the two-body escape velocity. See text for discussion.}
   \label{fig:water}
\end{center}
\end{figure}

In order to get an estimate on the loss of volatiles such as water ice on the surface of colliding bodies we track the overall water content of significant fragments after the collision. As initially all the water is present as surface ice on the target amounting to 30\,mass-\% the system's total water fraction is 0.15. Figure~\ref{fig:water} shows how this fraction develops between collision scenarios and reveals that in general more water is retained for ``less violent'' impacts---for $\alpha \lesssim 20^\circ\/$ and $v_0 \lesssim 1.3\,v_\mathrm{esc}\/$ almost all water ice stays on the survivor, for strongly inclined hit-and-go collisions most of the water stays as well. Generally, an increasing amount of water ice gets lost for smaller collision angles and higher velocities with a stronger dependency on velocity (impact energy) than on the angle.

Notable features in Fig.~\ref{fig:water} are the downward spikes occurring in the $v_0 \le 1.36\,v_\mathrm{esc}\/$ curves. These are due to a single surviving body that spins very quickly spraying debris into space and therefore losing large portions of its surface water ice. Again, we refer to fragment counts and individual scenario descriptions in \citet{maidvo13b} for more detailed discussions.

\section{Conclusions and further research}

We further evaluated results from an earlier study comparing the ``hydro'' and ``solid'' models for simulating small to mid scale collisions of planetesimals at moderate energies \citep{maidvo13b}. Rather than focusing on hydro-solid differences analyzing total fragment counts and three-biggest-fragment properties we were looking at the number of surviving ``significant fragments'' and their volatile content.  At a qualitative level the collision outcomes agree with analytic models for (giant) collisions of strengthless planets \citep[as given e.g., in ][]{leiste12} with some minor shifts of boundaries between the merging, erosion, and hit-and-run regimes. As expected there is more water ice surviving the collision for smaller velocities and more inclined impacts. Future studies will focus on (a) fragment dynamics determining their ability to escape the system's Hill sphere and (b) the fate of subsurface water ice inclusions as opposed to an icy shell.

{%\small
\begin{acknowledgements}
This research is produced as part of the FWF Austrian Science Fund project S~11603-N16.
%This publication was supported by FWF. 
In part the calculations for this work were performed on the hpc-bw-cluster---we gratefully thank the bwGRiD project\footnote{bwGRiD (http://www.bw-grid.de), member of the German D-Grid initiative, funded by the Ministry for Education and Research (Bundesministerium fuer Bildung und Forschung) and the Ministry for Science, Research and Arts Baden-Wuerttemberg (Ministerium fuer Wissenschaft, Forschung und Kunst Baden-Wuerttemberg).} for the computational resources.
\end{acknowledgements}
}

\bibliographystyle{aa} % style aa.bst
\bibliography{references} % references.bib

\end{document}